\documentclass[prl,twocolumn,superscriptaddress,showpacs,amscd,amsmath,amssymb,verbatim]{revtex4}

\usepackage{graphicx}
\usepackage[dvips]{hyperref}
\usepackage{textcomp}
\usepackage[OT1,T1]{fontenc}

\begin{document}

\newcommand{\herbert}{ZnCu$_{3}$(OH)$_{6}$Cl$_{2}$}
\newcommand{\kagome}{kagom\'e }

\title{Dzyaloshinsky-Moriya Anisotropy in the Spin-1/2 Kagom\'e Compound ZnCu$_{3}$(OH)$_{6}$Cl$_{2}$}
\author{A.~Zorko}
\affiliation{Laboratoire de Physique des Solides, UMR CNRS 8502, Universit\'e Paris-Sud 11, 91405 
Orsay, France}
\affiliation{Jo\v{z}ef Stefan Institute, Jamova 39, 1000 Ljubljana, Slovenia}
\author{S.~Nellutla}
\affiliation{National High Magnetic Field Laboratory, FSU, Tallahassee, FL, USA}
\author{J.~van~Tol}
\affiliation{National High Magnetic Field Laboratory, FSU, Tallahassee, FL, USA}
\author{L.~C.~Brunel}
\affiliation{National High Magnetic Field Laboratory, FSU, Tallahassee, FL, USA}
\author{F.~Bert}
\affiliation{Laboratoire de Physique des Solides, UMR CNRS 8502, Universit\'e Paris-Sud 11, 91405 
Orsay, France}
\author{F. Duc}
\affiliation{Centre d'\'Elaboration des Mat\'eriaux et d'\'Etudes Structurales, CNRS UPR 8011, 31055 Toulouse, France}
\author{J.-C.~Trombe}
\affiliation{Centre d'\'Elaboration des Mat\'eriaux et d'\'Etudes Structurales, CNRS UPR 8011, 31055 Toulouse, France}
\author{M.~A.~de~Vries}
\affiliation{School of Chemistry and SCEC, The University of Edinburgh, Edinburgh, EH9 3JZ, UK}
\author{A.~Harrison}
\affiliation{School of Chemistry and SCEC, The University of Edinburgh, Edinburgh, EH9 3JZ, UK}
\author{P.~Mendels}
\affiliation{Laboratoire de Physique des Solides, UMR CNRS 8502, Universit\'e Paris-Sud 11, 91405
Orsay, France}
\date{\today}
\begin{abstract}

We report the determination of the Dzyaloshinsky-Moriya interaction, the dominant magnetic anisotropy term in the \kagome spin-1/2 compound {\herbert}. Based on the analysis of the high-temperature electron spin resonance (ESR) spectra, we find its main component  $\left|D_z\right|=15(1)$~K to be perpendicular to the \kagome planes. Through the temperature dependent ESR line-width we observe a building up of nearest-neighbor spin-spin correlations below $\sim$150~K.

\end{abstract}
\maketitle

Among all geometrically frustrated networks, the spin-1/2 corner-sharing \kagome antiferromagnetic lattice has been at the forefront of the quest for novel quantum phenomena for the past two decades \cite{MisguichR}. Various competing ground states with minute energy difference have been proposed~\cite{MisguichR,kago} and are still under active consideration. The understanding of the ground state and of the low-lying excitations therefore still appears as a pending theoretical issue, even for the simplest nearest-neighbor (\textit{nn}) isotropic Heisenberg case. In addition, minor deviations from this model, often met in experimental realizations, such as magnetic anisotropy \cite{Elhajal, Ballou} and spinless defects \cite{Dommange}, may crucially affect the fundamental properties of the \kagome antiferromagnets.

\begin{figure}[b]
\includegraphics[width=7.3cm]{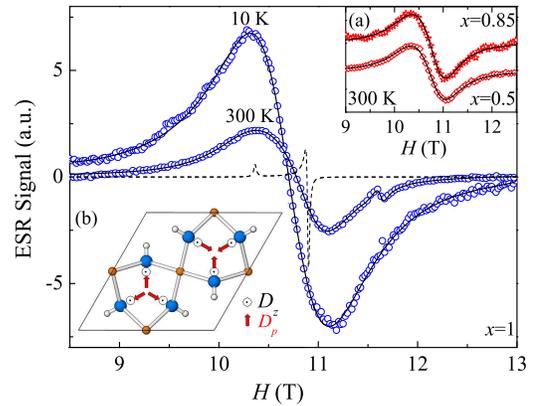}
\caption{(color online). ESR spectra of Zn$_{x}$Cu$_{4-x}$(OH)$_{6}$Cl$_{2}$, measured at 326.4~GHz, and 
their fits to the powder-averaged Lorentzian (solid lines). The dashed line gives an axial powder spectrum 
for the same $g$-factor anisotropy  ($g_\parallel=2.25$, $g_\perp=2.14$) in the case of small magnetic-anisotropy 
broadening. Inset (b): DM vectors connecting the \textit{nn} Cu$^{2+}$ cites (orange) through OH$^-$ groups (blue, gray), with $D_z$ and $D_p$  as the out-of-plane  and the in-plane components.}
\label{fig-1}
\end{figure}

In the pursuit of finding a suitable compound which would reflect intrinsic \kagome properties, 
the mineral herbertsmithite, \herbert, has recently been highlighted \cite{Shores} as the first "structurally perfect" 
realization of the spin-1/2 \kagome lattice [inset (b) to Fig.~\ref{fig-1}]. It features Cu$^{2+}$-based \kagome planes, 
separated by non-magnetic Zn$^{2+}$. It lacks any long-range magnetic order (LRO) down to at least 50~mK \cite{Mendels} 
despite sizable {\it nn} exchange $J \approx 190$~K \cite{Helton,Misguich}. Its bulk magnetic susceptibility $\chi_b$ 
monotonically increases with decreasing {\it T} \cite{Helton,Bert}, at variance with numerical calculations 
\cite{Elstner,Misguich,Rigol}. It was argued that this increase is due to magnetic anisotropy of the 
Dzyaloshinsky-Moriya (DM) type \cite{Rigol}, ${\bf D}_{ij}\cdot {{\bf S}_{i} \times {\bf S}_{j}}$, and/or due to 
impurities \cite{Misguich,Bert}. Various experiments indeed evidenced $4-7\%$ of Cu$^{2+}$/Zn$^{2+}$ inter-site 
disorder in the samples available at present \cite{Misguich,Bert,Olariu,deVries}. This explains why $\chi_b$ differs 
significantly from the local susceptibility of the \kagome planes $\chi_k$ measured by NMR \cite{Imai, Olariu}. 
The latter exhibits a pronounced decrease below $\sim${\it J}/2 and shows a finite plateau at low $T$ \cite{Olariu}. 
Together with ungapped excitations \cite{Olariu,Ofer,Imai} this points to the absence of the 
singlet-triplet gap in this compound.

The absence of the spin gap leads to the important question whether this is an intrinsic feature of the \kagome physics 
or is it related to additional terms in the Hamiltonian of {\herbert}. The presence and role of the DM interaction, highlighted by Rigol and Singh \cite{Rigol}, have been questioned on several occasions \cite{Misguich,Imai,Olariu}. Such perturbation to the isotropic Hamiltonian can indeed drastically affect the low-$T$ properties by mixing different states and can even stabilize 
LRO, as in jarosites \cite{Elhajal,Matan}. In order to be able to inspect this crucial issue it is of utmost importance to experimentally determine its magnitude in {\herbert}.

Using electron spin resonance (ESR), in this Letter we provide evidence of a sizable magnetic anisotropy, which we argue to be the DM interaction. In addition, through the temperature dependence of the line-width we show that spin correlations build up below $\sim$150~K and relate the low-T behavior to inter-site mixing defects.

Determined by the imaginary part of the dynamical susceptibility $\chi"(\omega)$, ESR spectra, $I(\omega)\propto \chi"({\rm \textbf{q}}\rightarrow0,\omega) \propto \int_{-\infty}^{\infty}{\rm d}t \left\langle S^+(t)S^-(0)\right\rangle \exp^{i \omega t}/T$, reflect the $T$-evolution of the spin correlation function $\left\langle S^+(t)S^-(0)\right\rangle$ in the direction perpendicular to the applied magnetic field $H$. Kramers-Kr\"onig relations demonstrate that the integrated ESR intensity $\chi_{\rm {\it ESR}}$ gives a locally measured static susceptibility. Further, a finite ESR line-width provides a direct measure of a finite magnetic anisotropy. 

\begin{figure}[t]
\includegraphics[width=6.6 cm]{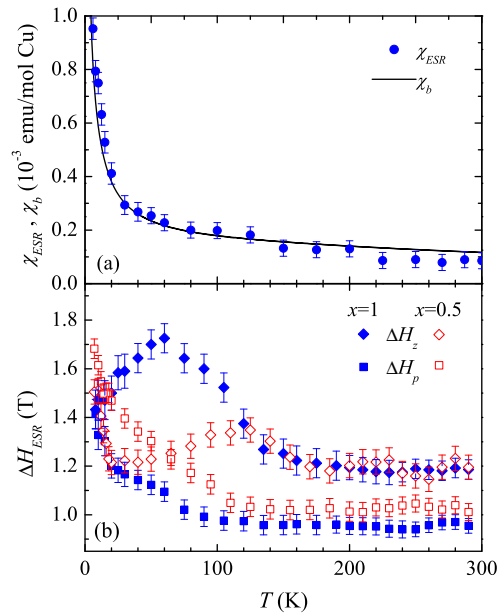}
\caption{(color online). (a) Scaling of ESR intensity $\chi_{\rm {\it ESR}}$ and bulk susceptibility $\chi_b$ in Zn$_{x}$Cu$_{4-x}$(OH)$_{6}$Cl$_{2}$ ($x=1$). (b) The out-of-plane $\Delta H_z$ and the in-plane 
$\Delta H_p$ ESR line-width.}
\label{fig-2}
\end{figure}

In Fig.~\ref{fig-1} we show typical derivative ESR spectra of polycrystalline Zn$_{x}$Cu$_{4-x}$(OH)$_{6}$Cl$_{2}$ ($x=0.5-1$) synthesized as in Ref.~\cite{Shores}, which were recorded at  326.4~GHz. Using a reference sample, in {\herbert} we find $\chi_{\rm {\it ESR}}=1.0(2)$~emu/molCu  at room temperature (RT), which is in good agreement with the bulk susceptibility 
$\chi_b=1.1$~emu/molCu. This proves that the Cu$^{2+}$ spins on the \kagome lattice are detected by ESR \cite{foot2}. The {\it T}-dependence of $\chi_{\rm {\it ESR}}$ is presented in Fig.~\ref{fig-2}(a). Its monotonic increase with decreasing {\it T} closely resembles $\chi_b$ down to 5~K and differs from the peaked \kagome planes susceptibility $\chi_k$ \cite{Imai,Olariu}. This demonstrates that ESR detects both the Cu$^{2+}$ spins on the \kagome lattice and those on the inter-layer Zn sites, the latter resulting from the Zn$^{2+}$/Cu$^{2+}$ inter-site disorder. Despite the two detected Cu$^{2+}$ sites with likely different $g$-factors ($\delta g \lesssim 0.2$), only a single ESR line is detected. This is expected \cite{Gulley} because the exchange coupling $J' \sim 10$~K between the two Cu$^{2+}$ species \cite{foot} is larger than their difference in Zeeman energy, $\delta g \mu_B H\lesssim 1$~K.

In order to fit the spectra appropriately, one has to consider both the $g$-factor anisotropy and the line-width broadening by the magnetic anisotropy. The former is given by $g(\theta)=(g_\parallel^2\cos^2(\theta-\theta_g)+ g_\perp^2\sin^2(\theta-\theta_g))^{1/2}$, where $\theta$ is the angle between $H$ and the normal $c$ to \kagome planes. $\theta_g=$36{\textdegree} denotes the tilt of CuO$_4$ plaquettes with respect to the \kagome planes \cite{Shores}, which locally sets the principal axes of the $g$-tensor. Typical $g$-factor values for Cu$^{2+}$ ions in a uniaxial crystal field $g_\parallel=2.2-2.4$ and $g_\perp=2.05-2.15$ should yield two distinctive narrow features in a single ESR derivative spectrum (Fig.~\ref{fig-1}). In contrast, the observed spectra appear as a single {\it broad} derivative shape, which is a direct evidence of a {\it sizable} magnetic anisotropy. In the case of the DM interaction, which we argue below to be dominant, the symmetry of the lattice imposes the line-width to depend only on $\theta$. To fit the spectra, we therefore convoluted a powder field distribution and a Lorentzian, the latter having the semi-phenomenological line-width $\Delta H = (\Delta H_z^2\cos^2\theta+\Delta H_p^2\sin^2\theta)^{1/2}$, where $\Delta H_z$ and $\Delta H_p$ correspond to the direction parallel and perpendicular to $c$. The field distribution combines a uniform angular distribution and the $g$-factor anisotropy $g(\theta)$. Standard $g$-tensor values, $g_\parallel=2.25$ and $g_\perp=2.14$, were found.
We stress that the RT ESR spectra are not noticeably affected by the Cu$^{2+}$ spins residing on the Zn sites. They can be fitted with the same set of parameters for $x=0.5-1$ (Fig.~\ref{fig-1}). We therefore use them later to determine the magnetic anisotropy in the \kagome planes.

Several distinct regimes are observed in the $T$-dependence of the ESR line-width [Fig.~\ref{fig-2}(b)]. It is constant above $\sim$200~K, with very similar values in {\herbert} and Zn$_{0.5}$Cu$_{3.5}$(OH)$_{6}$Cl$_{2}$. This $T$-independence is characteristic of the high-$T$ exchange narrowing regime. Below $\sim$150~K a pronounced increase is observed in $\Delta H_z$, while $\Delta H_p$ starts increasing noticeably below $\sim$100~K. On general grounds \cite{Richards}, we recognize the increase of the line-width as the evidence of a building up of {\it short-range spin correlations} at $T\lesssim J$. The difference between $\Delta H_z$ and $\Delta H_p$ indicates that the correlations are evolving in an anisotropic manner. We propose that this results from magnetic anisotropy affecting the evolution of the correlations. With decreasing temperature the ESR spectra should  broaden monotonically as the spin correlations develop, e.g., similarly as found in another two dimensional (2D) frustrated system, SrCu$_2$(BO$_3$)$_2$ \cite{Zorko}. At variance, a maximum is observed in $\Delta H_z$. Its position shifts significantly toward higher temperatures with reducing the Zn content [Fig.~\ref{fig-2}(b)]. We therefore relate this maximum to the Zn$^{2+}$/Cu$^{2+}$ inter-site disorder. The effect of different Cu$^{2+}$ spins on the ESR signal is proportional to their susceptibility \cite{Gulley}. Therefore, below $\sim$20~K where $\chi_b$ overshadows $\chi_k$, the Cu$^{2+}$ spins on the Zn sites dominate the ESR response, which explains the crossover between 20~K and 150~K. Weakly coupled Cu$^{2+}$ spins on the Zn sites, with a different environment, hence a different magnetic anisotropy, would provide a natural explanation of the low-$T$ behavior. Substantial theoretical modeling is needed though to account quantitatively for the observed line-width $T$-dependence, which is beyond the scope of this paper.

We address the origin and magnitude of the magnetic anisotropy in {\herbert} from the detailed analysis of its RT spectrum, in the framework of the Hamiltonian

\begin{equation}
\label{eq1}
\mathcal{H}= J \sum _{(ij)} {{\bf S}_{i} \cdot {\bf S}_{j}} - g\mu_{B}H \sum_{i} S_i^z + \mathcal{H'},
\end{equation}

\noindent where, the first sum runs oven the {\it nn} spin pairs and represents the exchange interaction $\mathcal{H}_e$, the second one gives Zeeman 
coupling $\mathcal{H}_Z$, and $\mathcal{H}'$ is magnetic anisotropy. For strong exchange with respect to $\mathcal{H}_Z$ and $\mathcal{H}'$ the Kubo-Tomita (KT) formalism \cite{Kubo} yields a Lorentzian ESR spectrum, in agreement with our experiment. Regardless of the specific form of the anisotropy, its line-width can be estimated from the magnitude $A$ of the anisotropy as $\Delta H \sim A^2/J$ \cite{Castner}, yielding $A \sim 16$~K. 

We stress again that the spectra are extremely broad. Both, the dipolar interaction between Cu$^{2+}$ spins and their hyperfine coupling to nuclear spins are too small to account for the observed line-width. They would lead to $\Delta H\lesssim 1$~mT due to strong exchange narrowing. Therefore, exchange anisotropy must be at work in {\herbert}. It originates from a sizable spin-orbit coupling $\lambda$, which mixes orbital excited states of Cu$^{2+}$ with their Kramer's ground doublet, and is reflected in measured $g$-factor shifts of 0.14-0.25 from the free electron value 2.0023. The second-order perturbation calculation yields two exchange anisotropy terms -- antisymmetric DM interaction $D\propto(\Delta g/g)J$ (linear in $\lambda$) and symmetric anisotropic exchange (AE) $\Gamma\propto(\Delta g/g)^2J$ (quadratic in $\lambda$) \cite{Moriya}. The DM interaction $\mathcal{H}'=\sum_{(ij)}{\bf D}_{ij}\cdot {{\bf S}_{i} \times {\bf S}_{j}}$ is thus usually dominant in Cu-based antiferromagnets, if allowed by symmetry \cite{Moriya}. 

The out-of-plane DM component $D_z$ is generally present in the \kagome lattice, while the in-plane component $D_p$ is allowed in {\herbert} because the superexchange mediating O$^{2-}$ ions break the mirror symmetry of the \kagome planes. We therefore base our ESR analysis on the {\it dominance of the DM anisotropy}.  The ${\bf D}_{ij}$ pattern (Fig.~\ref{fig-1}) is obtained by choosing the direction of one of the vectors and applying symmetry operators of the lattice space group. We use the convention of spins being counted clockwise in all triangles.

We can calculate the ESR line-width \cite{foot3}

\begin{align}
\Delta H(\theta) &= \sqrt{2 \pi} \frac{k_b}{2g(\theta) \mu_B J}\notag \\%
\label{eq3}%
&\sqrt{ \frac{\left(2D_z^2+3D_p^2+ \left(2D_z^2-D_p^2 \right)\cos^2\theta \right)^3}{16D_z^2+78D_p^2+\left( 16D_z^2-26D_p^2 \right)\cos^2\theta}},
\end{align}

\begin{figure}[t,b]
\includegraphics[width=7.3cm]{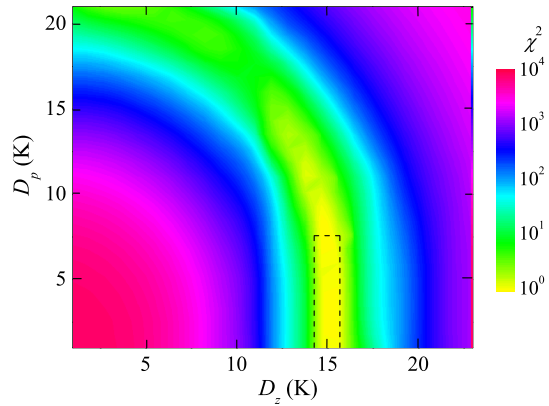}
\caption{(color online). The reduced $\chi^2$, reflecting the quality of the RT fit in {\herbert}. The dashed rectangle gives the optimal parameters $\left|D_z\right|=15(1)$~K and $\left|D_p\right|=2(5)$~K.}
\label{fig-3}
\end{figure}

\noindent determined by the DM pattern (Fig.~\ref{fig-1}), in the infinite-$T$ limit. The high-$T$ exchange narrowing 
regime is reached at RT since the ESR spectra do not change noticeably above $\sim$200~K. This is consistent 
with a $T$-independent NMR spin-lattice relaxation above 150~K \cite{Imai}. We fitted the RT ESR spectrum with the 
powder-averaged Lorentzian having the above line-width $\Delta H(\theta)$. In Fig.~\ref{fig-3} we plot on the 
$D_z$-$D_p$ map the reduced $\chi^2$ obtained from the fit. Optimal parameters are $\left|D_z\right|=15(1)$~K and $\left|D_p\right|=2(5)$~K. According to Eq.~(\ref{eq3}) their sign cannot be determined by ESR. We find the magnitude of the extracted DM interaction ($D/J \approx 0.08$) in the range set by several other Cu-based 2D frustrated systems. For instance in the orthogonal-dimer system SrCu$_2$(BO$_3$)$_2$ it amounts to 4\% of the isotropic exchange \cite{Zorko,Cheng}, while the ratio of 16\% was reported for the triangular compound Cs$_2$CuCl$_4$ \cite{Coldea}.

At this point it is important to address the issue of the possible hidden symmetry of the DM interaction \cite{KaplanShekhtman}. For linear spin chains with staggered DM vectors one can effectively transform it to the AE term of the magnitude $D^2/J$ by applying a non-uniform spin rotation \cite{Choukroun}. This makes it of the same order as the initial AE and discards the direct applicability of the KT formalism \cite{Choukroun,Oshikawa}. However, for other lattices one should distinguish between reducible DM components, transforming to $\lambda ^2$ order, and irreducible components, remaining linear in $\lambda$ \cite{Cheng}. The components that can be eliminated in the first order are those which sum up to zero within any closed loop on the lattice \cite{Cheng}. For the \kagome lattice, the in-plane $D_p$ is reducible while the out-of-plane $D_z$ is irreducible. Using the KT formalism for the latter is therefore well justified, while the former might not be accurate. However, the small value of $D_p$ does not significantly influence the ESR spectra (see Fig.~\ref{fig-3}). We note that in SrCu$_2$(BO$_3$)$_2$ the values of the DM interaction obtained from the KT analysis \cite{Zorko}, from inelastic neutron scattering \cite{Cheng} and from NMR measurements \cite{Kodama} were found to agree within 20\%. 

In Ref. \cite{Rigol} somewhat larger values, $\left|D_z\right|\approx 0.1J$ and $\left|D_p\right|\approx 0.2-0.3J$, than extracted in this study were suggested to explain the bulk susceptibility upturn in {\herbert}, by considering the DM interaction as the only additional term to the Heisenberg Hamiltonian. These values would lead to high-$T$ ESR line-widths of $6-7$~T, which is inconsistent with our experimental observations. Therefore, DM interaction alone cannot account for the macroscopic susceptibility in {\herbert}.

Sizable DM interaction ($D/J \approx 0.08$) raises the fundamental question {\it to what extent DM anisotropy affects 
the low-{\it T} magnetism of} {\herbert}. Spin-wave excitations in the spin-5/2 Fe-jarosite were explained by the 
presence of the DM interaction of a similar magnitude ($D/J \approx 0.09$) \cite{Matan}. In this  \kagome compound, N\'eel order 
below 65~K was reproduced by classical Monte-Carlo calculations, predicting uniform ($\textbf{q}=0$) 
ordered spin structures induced by the DM interaction \cite{Elhajal}. In the $S=1/2$ case, quantum correction should suppress LRO 
for small $D/J$ \cite{Ballou}. For $D/J \gtrsim 0.05$ recent numerical calculations seem to indicate a broken-symmetry 
state at $T=0$, which however has no on-site magnetic moment \cite{Fong}.
Further, the rather large Cu$^{2+}$/Zn$^{2+}$ inter-site disorder may destabilize LRO. Indeed, vacancies on the spin-1/2 Heisenberg \kagome lattice induce unexpectedly extended dimer-dimer correlations, despite short-ranged spin-spin correlations \cite{Dommange}. Although exact dimer patterns are not known for additional terms in the Hamiltonian that break the SU(2) symmetry, it seems plausible that they would contradict the uniform LRO state preferred by the DM interaction. This might explain the absence of LRO in {\herbert} down to $D/300=50$~mK.

Last, quite importantly, the DM interaction mixes magnetic excitations into the presumably singlet ground state of the 
Heisenberg \kagome lattice. This could explain the experimentally observed gapless excitation spectrum in {\herbert} 
\cite{Olariu,Imai} and the finite susceptibility of \kagome planes at low {\it T} \cite{Olariu}, as
confirmed by numerical calculations on finite spin clusters \cite{Samir}. These experimental results could likewise be described as a disorder-generated effect, due to a vacancy-induced spatially distributed density of states \cite{Gregor}. Most likely, both effects play an important role.

In conclusion, we have shown that the DM interaction in {\herbert} is sizable ($D/J \approx 0.08$) and could critically affect the low-{\it T} properties. Knowing its magnitude, future theoretical studies should investigate its impact on the ground state. They should also clarify whether the DM magnetic anisotropy and the Cu$^{2+}$/Zn$^{2+}$ inter-site disorder are primarily responsible for the absence of the spin gap.

We acknowledge discussions with B. Canals, O. C\'epas, C. Lacroix, C. Lhuillier, G. Misguich and S. El Shawish. 
The work was supported by the EC MC Grant MEIF-CT-2006-041243, the ANR project NT05-4\_41913 and 
the ARRS project Z1-9530. NHMFL is supported by NSF (DMR-0654118) and the State of Florida.

\appendix


\begin{thebibliography}{000}
%
\bibitem{MisguichR} For a review see G.~Misguich and C.~Lhuillier, {\it Frustrated Spin Systems}, ed. by H. T. Diep (World-Scientific, 2005).\label{MisguichR}
%
\bibitem{kago} Y.~Ran, {\it et al.},
Phys. Rev. Lett. \textbf{98}, 117205 (2007); S.~Ryu {\it et al.}, Phys. Rev. B \textbf{75}, 184406 (2007); R.~R.~P.~Singh and D. A. Huse, Phys. Rev. B \textbf{76}, 180407(R) (2007).\label{kago}
%
\bibitem{Elhajal} M.~Elhajal {\it et al.},
Phys. Rev. B \textbf{66}, 014422 (2002).\label{Elhajal}
%
\bibitem{Ballou} R.~Ballou {\it et al.},
J. Mag. Mag. Mat. \textbf{262}, 465 (2003).\label{Ballou}
%
\bibitem{Dommange} S.~Dommange  {\it et al.},
Phys. Rev. B \textbf{68}, 224416 (2003).\label{Dommange}
%
\bibitem{Shores} M.~P.~Shores {\it et al.}, J. Am. Chem. Soc. \textbf{127}, 13462 (2005).\label{Shores}
%
\bibitem{Mendels} P.~Mendels {\it et al.},
Phys. Rev. Lett. \textbf{98}, 077204 (2007).\label{Mendels}
%
\bibitem{Helton} J.~S.~Helton {\it et al.},
Phys. Rev. Lett. \textbf{98}, 107204 (2007).\label{Helton}
%
\bibitem{Misguich} G.~Misguich and P.~Sindzingre,
Eur. Phys. J. B \textbf{59}, 305 (2007).\label{Misguich}
%
\bibitem{Bert} F.~Bert {\it et al.},
Phys. Rev. B \textbf{76}, 132411 (2007).\label{Bert}
%
\bibitem{Elstner} N.~Elstner and A.~P.~Young,
Phys. Rev. B \textbf{50}, 6871 (1994).\label{Estner}
%
\bibitem{Rigol} M.~Rigol and R.~R.~P.~Singh,
Phys. Rev. Lett. \textbf{98}, 207204 (2007); Phys. Rev. B \textbf{76}, 184403 (2007).\label{Rigol}
%
\bibitem{Olariu} A.~Olariu {\it et al.},
Phys. Rev. Lett. \textbf{100}, 087202 (2008).\label{Olariu}
%
\bibitem{deVries} M.~A.~de Vries {\it et al.}, Phys. Rev. Lett. \textbf{100}, 157205 (2008).\label{deVries}
%
\bibitem{Ofer} O.~Ofer {\it et al.}, arXiv:cond-mat/0610540.\label{Ofer}
%
\bibitem{Imai} T.~Imai {\it et al.},
Phys. Rev. Lett. \textbf{100}, 077203 (2008).\label{Imai}
%
\bibitem{Matan} K.~Matan {\it et al.},
Phys. Rev. Lett. \textbf{96}, 247201 (2006).\label{Matan}
%
\bibitem{foot2} The intensity of the small impurity-phase signal observed around 11.6~T is ca. 0.2\% of the total ESR intensity.\label{foot2}
%
\bibitem{Gulley} J.~E.~Gulley and V.~Jaccarino,
Phys. Rev. B \textbf{6}, 58 (1972).\label{Gulley}
%
\bibitem{foot} This order of magnitude is inferred from the magnetic ordering temperature (6-19~K) of the Zn$_{x}$Cu$_{4-x}$(OH)$_{6}$Cl$_{2}$ compounds with $x<1$ \cite{Shores,Mendels}.\label{foot}
%
\bibitem{Richards} P.~M.~Richards, {\itshape Local Properties of
Low-Dimensional Antiferromagnets}, ed. by K. A. Müller (Nord Holland Publishing Company, Amsterdam, 1976).\label{Richards}
%
\bibitem{Zorko} A.~Zorko {\it et al.},
Phys. Rev. B \textbf{69}, 174420 (2004).\label{Zorko}
%
\bibitem{Kubo} R.~Kubo and K.~Tomita, J. Phys. Soc. Jpn. \textbf{9}, 888 (1954).\label{Kubo}
%
\bibitem{Castner} T.~G. Castner,~Jr. and M.~S.~Seehra, Phys. Rev.
B \textbf{4}, 38 (1971).\label{Castner}
%
\bibitem{Moriya} T.~Moriya,
Phys. Rev. \textbf{120}, 91 (1960).\label{Moriya}
%
\bibitem{foot3} The ESR line-width $\Delta H = \sqrt{2\pi} k_{B}/(g \mu _{B}) \sqrt{M_{2}^{3}/M_{4}}$
is for $k_B T \gg g\mu_{B}H$ given by the second and the fourth moment, $M_{2}=\left\langle \left[ \mathcal{H}^{\prime },S^{+}\right]
[ S^{-},\mathcal{H}^{\prime }] \right\rangle /\left\langle S^{+}S^{-}\right\rangle$ 
and
$M_{4}=\left\langle \left[ \mathcal{H}-\mathcal{H}_{Z},\left[
\mathcal{H}^{\prime },S^{+}\right] \right] [ \mathcal{H}-\mathcal{H}_{Z},\left[
\mathcal{H}^{\prime },S^{-}\right] ] \right\rangle / \left\langle
S^{+}S^{-}\right\rangle$ \cite{Castner}.\label{foot3}
%
\bibitem{Cheng} Y.~F.~Cheng {\it et al.},
Phys. Rev. B \textbf{75}, 144422 (2007).\label{Cheng}
%
\bibitem{Coldea} R.~Coldea {\it et al.},
Phys. Rev. Lett. \textbf{88}, 137203 (2002).\label{Coldea}
%
\bibitem{KaplanShekhtman} T.~A.~Kaplan,
Z. Phys. B: Condens. Matter \textbf{49}, 313 (1983); L. Shekhtman, O. Entin-Wohlman, and A. Aharony, Phys. Rev. Lett. \textbf{69}, 836 (1992).\label{KaplanShekhtman}
%
\bibitem{Choukroun} J.~Choukroun {\it et al.},
Phys. Rev. Lett. \textbf{87}, 127207 (2001).\label{Choukroun}
%
\bibitem{Oshikawa} M.~Oshikawa and I.~Affleck,
Phys. Rev. B \textbf{65}, 134410 (2002).\label{Oshikawa}
%
\bibitem{Kodama} K.~Kodama {\it et al.}, J. Phys.: Condens. Matter
\textbf{17}, L61 (2005).\label{Kodama}
%
\bibitem{Fong} O.~C\'epas {\it et al.}, arXiv:cond-mat/0806.0393.
\label{Fong}

\bibitem{Samir} S.~El~Shawish, private communications
(unpublished).\label{Samir}
%
\bibitem{Gregor} K.~Gregor and O.~I.~Motrunich, Phys. Rev. B \textbf{77}, 184423 (2008).\label{Gregor}

\end{thebibliography}
\end{document}